\documentclass{bmcart}

\usepackage{amsthm,amsmath}
\usepackage[utf8]{inputenc} 

\startlocaldefs

\usepackage{soul}
\usepackage{tabularx,graphicx}
\usepackage{enumerate}
\usepackage{hyperref}

\endlocaldefs

\begin{document}

\begin{frontmatter}

\begin{fmbox}
\dochead{Research}


\title{Internal migration and mobile communication patterns among pairs with strong ties}


\author[
   addressref={apctp,uvm},
]{\inits{MIDF}\fnm{Mikaela Irene D.} \snm{Fudolig}}
\author[
   addressref={aalto_cs},
]{\inits{DM}\fnm{Daniel} \snm{Monsivais}}
\author[
   addressref={aalto_iem,aalto_cs},
]{\inits{KB}\fnm{Kunal} \snm{Bhattacharya}}
\author[
   addressref={cath_univ},
   email={h2jo@catholic.ac.kr}  
]{\inits{HHJ}\fnm{Hang-Hyun} \snm{Jo}}
\author[
   addressref={aalto_cs,turing},
]{\inits{KK}\fnm{Kimmo} \snm{Kaski}}


\address[id=apctp]{
  \orgname{Asia Pacific Center for Theoretical Physics}, 
  \city{Pohang},                              
  \cny{Republic of Korea}                                    
}

\address[id=uvm]{
  \orgname{Vermont Complex Systems Center, University of Vermont}, 
  \city{Burlington, Vermont},                              
  \cny{USA}                                    
}

\address[id=aalto_cs]{
  \orgname{Department of Computer Science, Aalto University School of Science}, 
  \city{Espoo},                              
  \cny{Finland}                                    
}

\address[id=aalto_iem]{
  \orgname{Department of Industrial Engineering and Management, Aalto University School of Science}, 
  \city{Espoo},                              
  \cny{Finland}                                    
}

\address[id=cath_univ]{
  \orgname{Department of Physics, The Catholic University of Korea}, 
  \city{Bucheon},                              
  \cny{Republic of Korea}                                    
}

\address[id=turing]{
  \orgname{The Alan Turing Institute}, 
  \city{London},                              
  \cny{United Kingdom}                                    
}




\end{fmbox}


\begin{abstractbox}

\begin{abstract} 
\label{sec:abstract}

\noindent Using large-scale call detail records of anonymised mobile phone service subscribers with demographic and location information, we investigate how a long-distance residential move within the country affects the mobile communication patterns between an ego who moved and a frequently called alter who did not move. By using clustering methods in analysing the call frequency time series, we find that such ego-alter pairs are grouped into two clusters, those with the call frequency increasing and those with the call frequency decreasing after the move of the ego. This indicates that such residential moves are correlated with a change in the communication pattern soon after moving. We find that the pre-move calling behaviour is a relevant predictor for the post-move calling behaviour. While demographic and location information can help in predicting whether the call frequency will rise or decay, they are not relevant in predicting the actual call frequency volume. We also note that at four months after the move, most of these close pairs maintain contact, even if the call frequency is decreased.

\end{abstract}


\begin{keyword}
\kwd{mobile phone data}
\kwd{call detail records}
\kwd{migration}
\kwd{residential mobility}
\kwd{communication patterns}
\end{keyword}


\end{abstractbox}
%

\end{frontmatter}



\section*{Introduction}
\label{sec:intro}

In recent years, the availability of large-scale mobile phone call detail records (CDRs) with location information has allowed researchers to study human mobility and communication with fine spatiotemporal resolution, large sample sizes, and high accuracy~\cite{blondel_survey_2015,bhattacharya_social_2019}. This has generated a lot of interest in both the individual mobility patterns in relation to activity spaces~\cite{jo_spatiotemporal_2012,jarv_understanding_2014,puura_relationship_2018} as well as in general patterns of human mobility~\cite{gonzalez_understanding_2008,schneider_unravelling_2013,palchykov_inferring_2014}.

Most of today's research on human mobility that use the mobile phone traces of individuals is focused on short-term daily mobility. On the other hand, the effects of moving to another location for longer term have not been studied as much, perhaps due to the lack of mobile phone datasets that allow the tracking of individuals' locations over long periods of time. These residential moves, referred to as migration or residential mobility in sociological literature~\cite{magdol_people_2000,coulter_re-thinking_2016}, have conventionally been studied using surveys and censuses. In contrast to these traditional approaches, mobile phone CDRs do not suffer from recall bias, show more subtle migration patterns~\cite{blumenstock_inferring_2012}, and give migration estimates as accurate as those obtained from census data~\cite{lai_exploring_2019}.

Although the migration flows are usually focused on individuals, migration also carries a relational aspect~\cite{findlay_new_2015}, and has been associated with changes in the individual's social network~\cite{shklovski_residential_2006,oishi_residential_2013,viry_residential_2012}. These changes have traditionally been studied using surveys~\cite{magdol_people_2000,lubbers_longitudinal_2010}, but the availability of the mobile phone CDRs have broadened the research scope to such issues as the persistence of ties in a mover's social network~\cite{lambiotte_geographical_2008,onnela_geographic_2011}. There it was found that the geographic distance between an ego and an alter has a diminishing effect on their mobile phone communication. Moreover, it has been shown that on the basis of communication volumes between ego-alter pairs, stronger ego-alter ties are more likely to persist after migration~\cite{phithakkitnukoon_out_2011}.

Most studies on migration involving CDRs take an egocentric approach in the study of migration, focusing on the mover~\cite{hankaew_inferring_2019} and its communication patterns~\cite{yang_stay_2018,hong_characterization_2019}. In contrast, there is a lack of studies involving mobile phone data that take into account the relational aspects of migration. Although the CDRs have a high spatiotemporal resolution and no recall bias, they usually do not contain the relationship information between users. Still, we can make some inferences. The regularity and frequency of mobile communication can be taken to indicate the strength of social ties~\cite{phithakkitnukoon_mobile_2010,saramaki_persistence_2014,roberts_communication_2011}, while the age and gender difference between a pair of close users can help infer the type of relationship between them~\cite{david-barrett_communication_2016,fudolig_different_2020,fudolig_link-centric_2020}. For example, an age difference of 20--40 years between a pair of regularly communicating users more likely corresponds to a parent-child pair than a pair of spouses. As our dataset contains information on mobile communication patterns, location information, and demographic information, we are in a unique position to study the interplay between the internal migration and the mobile communication patterns of ego-alter pairs from a relational perspective.

In this study, we look at the evolution of strong ties between the movers and their non-mover contacts in the case of long-distance residential moves. In particular, we focus on how the mobile communication frequency changes after the move and how this relates to the demographics and location information of both the mover ego and the non-mover alter. We approach our research question from a predictive standpoint: if we can predict whether movers will continue to contact close alters based on easily available information, decay of social ties may be prevented by proper intervention. Aside from the social benefits, this also has business advantages to the service provider as this encourages use of the service. Although our study focuses on CDRs, the same approach may be used for messaging apps for the same reasons.

In the present study, we use a dataset containing the CDRs of over a million mobile phone service subscribers, leveraging the fine spatiotemporal resolution, completeness of mobile communication network information, and large sample sizes that are typically lacking in surveys. We begin by inferring the home locations of the users and identifying the strong ego-alter ties in the mobile communication network. We then investigate both the number of calls and the fraction of calls dedicated by the mover ego to the non-mover alter, and use clustering methods to get insight into changes in the calling patterns between the ego and alter. Finally we study whether the change in the calling patterns can be predicted by pre-move calling behaviour, demographics, and location information.

\section*{Methodology}
\label{sec:method}

\subsection*{The dataset}

The dataset we use contains anonymised CDRs from a European country collected over a two year period from January 2008 to December 2009. The CDRs contain the following information: origin (ego) user ID, destination (alter) user ID, type of interaction (either call or SMS), date and time of interaction, an incoming/outgoing marker, and the origin user's cell tower ID. In a separate file, we also have a list of over 1.8 million user IDs active from 2008 to 2009 and their associated ages and genders in 2008 which were obtained from the users' contract information. In case of multiple users in a family contract ID that is represented by a single age and gender, only the user with the earliest contract start date is assigned the demographic information, while the other users are considered to have unknown age and gender. In addition, we have yet another file containing cell tower IDs and their geographical latitudes and longitudes. Note that not all the user IDs and cell tower IDs appear both in the CDRs and in the latter two files.

\subsection*{Identifying the movers}

Let us first consider the choice of spatial and temporal resolution for the analysis. Here we will focus on individual users' long-distance residential moves at the province level, which is the administrative level above the city or municipality level. This is based on the notion that although high-volume and affordable commuting transportation modes like the metro often extend from major cities to neighbouring towns, they rarely extend from one province to another, so changing location at the province level likely indicates a residential move instead of a home-work commute. We also choose the temporal resolution to be one month, as this resulted in reduced noise but still contained information enough to infer the home location of the user. Further, the monthly resolution allows us to see long-distance moves that were not abrupt (e.g., a person may repeatedly travel to and from the old and new home locations before moving to look for apartments, or after moving to transport furniture), which may be missed if we look for moves at finer temporal resolutions.

In order to find the home location of the user, we use Nominatim~\cite{openstreetmap_foundation_osm-searchnominatim_nodate} and the Python module \texttt{geopy} to find the address corresponding to the cell tower ID of every outgoing call and SMS of the user. If a cell tower location is unknown, the address is marked \texttt{NaN}. For each day, we take the most frequently recorded province among all entries including those marked as \texttt{NaN}. We call this the \textit{daily most common location} at the province level.

Although most studies use the night locations to infer the home locations~\cite{jo_spatiotemporal_2012,phithakkitnukoon_out_2011,hoteit_filling_2016}, we do not impose such a time limit. Since inter-province transportation is not as cheap and accessible as intra-province transportation, users are less likely to commute daily across provinces. Thus, we do not expect regular differences between the day or night locations at the province level. Relaxing the time constraint also increases the viable sample size (albeit at a coarser spatial resolution), as there are more records to base the home location on and thus fewer users with unidentified home locations. In addition, we have found that if months with unknown home locations are excluded from the analysis, the province-level trajectories based on night-time calls and SMS are mostly consistent with those obtained using all-day calls and SMS. Thus, the all-day phone activity provides comparable reliability to the night-time phone activity in obtaining home locations at the province level (see the Supplementary Information for more details on using night-time home locations). We then took the most common location among the daily most common locations during weekdays for every month, when people are expected to be in their primary residence, and used this as the home location. In case of ties, one location is chosen at random. Each user will then have a home location vector with 24 entries, each corresponding to a month in the observation period. For a comparison of our home detection method with others in the literature, we refer the reader to Ref.~\cite{chi_general_2020}.

Here we define ``movers'' as users who move from one home location, \texttt{province\_1}, to another home location, \texttt{province\_2}, only once within the 24-month period such that itineraries like (\texttt{province\_1 $\rightarrow$ province\_2 $\rightarrow$ province\_1})  or (\texttt{province\_1 $\rightarrow$ province\_2 $\rightarrow$ province\_3}) are not included in the analysis. Users with the run-length encoded home location vectors [(\texttt{province\_1}, $m$), (\texttt{province\_2}, $24-m$)] were considered  movers, where the provinces are known and $m$ is the number of months spent in \texttt{province\_1}. Thus, $m$ can be considered as the \textit{estimated moving month}, where $m\in[1,24]$ with $m=1$ representing January 2008. Users with home locations given by [(\texttt{province\_1}, 24)] are considered ``non-movers'', while those with other run-length encodings are considered to have unknown trajectories and are classified as neither movers nor non-movers. We also require the movers to stay in each home location for a certain number of months as described in the next subsection. In our dataset, we found 13,611 movers and 969,513 non-movers, while 843,333 users could not be categorized in absolute terms (see Supplementary Information Table S1 for details). The vast majority of the users who were not classified as either movers or non-movers have at least one month with an unknown home location or have two home locations but moved back and forth, possibly for short-term vacations. We note that the fraction of movers among movers and non-movers is 1.3\%, which is reasonable given the migration statistics of the country where the data was collected. To compare, the total number of interprovincial migratory flows from 2008--2009, which also includes short-term movers, is around 2\% of the country's population according to the census data.

\subsection*{Identifying strong ties}

We identify a mover ego's strong ties prior to the move by looking for both the frequency and regularity of mobile communication between the ego and its alters, similar to what we have done in our previous research on CDRs~\cite{fudolig_different_2020,fudolig_link-centric_2020}. For each ego and each month, the alters are ranked based on the fraction of the calls made or received from the ego that are associated with the alter, which we consider as a proxy for the relative volume of mobile communication devoted by the ego to the alter. If $c_i$ is the number of calls between the ego and one of its $n$ alters $i$, the fraction of calls to the alter $i$ is given by $c_i/\sum_{j=1}^n c_j$. The alter $i$ with the highest fraction is given rank 1; alters with tied fractions are given the same rank. Note that in the ranking procedure, we do not yet take into account the demographic and location information of the alters.

We limit our interest to pairs in which the ego moved and the alters did not move. The non-mover alters who were consistently ranked in the top five of the mover ego at least for months $m-2$, $m-1$, and $m$, where $m$ is the estimated moving month, are considered to have strong ties with the ego before moving. Among these ego-alter pairs, we examine those in which the ego and alter (1) have known age and gender, (2) had made or received at least one call or SMS from four months before the moving month to four months after (i.e. $m-4$ to $m+4$), and (3) have no unknown home locations. These filtering methods yield 4,487 ego-alter pairs, with 3,661 unique mover egos who lived in each of its home locations for at least four months and 4,453 unique non-mover alters. As there are a number of pairs with shared egos, we perform a train-test split on the unique egos and assign pairs to the set where the mover ego belongs. As this scheme results in no common egos between the train and test set, it minimizes the chance that the models learn the behavior from an ego in the train set and apply it to the same ego in the test set.

\subsection*{Examining changes in the call frequency patterns}

Although we have data on the SMS of the users, we focus on the call frequency for two main reasons. Firstly, as calls require immediate feedback from both the caller and the callee, we can assume each call as a unit of exchange; the more calls there are, the more communication there is between the two individuals. Secondly, since a single message can be sent as a single SMS or broken down into several SMS, and no immediate feedback is required on the part of the receiver, it is more difficult to infer communication volume using SMS. Previous studies on the same dataset show that including SMS messages does not significantly increase efficacy in ranking the alters~\cite{palchykov_close_2013}.

For each ego-alter pair, we construct the time series of two quantities measured at a monthly resolution: (1) the number of calls exchanged by the mover ego and the non-mover alter, and (2) the fraction of the calls made and received by the mover ego that are associated with the non-mover alter. The number of calls gives an estimate of the mobile communication frequency between the ego and the alter, while the fraction of calls is a measure of the relative importance of an alter compared to the other alters. We measure the time in terms of how far away a month is from the moving month: the moving month is taken to be $t=0$, while months before moving have values $t<0$ and months after moving have values $t>0$. To examine whether there is a change in the call frequency associated with a residential move, we use clustering methods on the time series for the different pairs. While it is possible to analyse this by separating time into pre-move and post-move periods and comparing them, exploratory clustering does not have restrictions on when changes are expected and thus gives a better qualitative picture on how calling behaviour changes in relation to a residential move.

Since we are interested more in the call frequency change rather than the actual call frequency value, we standardise each time series to mean 0 and standard deviation 1 over the time period of interest, i.e., $t\in[-4,4]$. This period was selected to include a sufficiently long time after moving without compromising sample size; further, since we only required that the alter be in the ego's top five for $t\in[-2,0]$, there are some pairs where the alter is a recent introduction to the ego's close circle, although such cases composed a very small minority of all the pairs considered (0.9\%). Since we are interested not just in the shape of the time series but also in the time at which changes occur, we do not use shape-preserving clustering methods~\cite{aghabozorgi_time-series_2015}. We use \textit{k}-means as our clustering method and use the mean silhouette score, the Davies-Bouldin index, and the Jaccard bootstrap similarity index~\cite{yu_bootstrapping_2019} to find the optimal number of clusters and quantify the quality of the clustering. Similar results were also obtained using hierarchical clustering with Ward's method.

In order to verify whether the resulting clusters are genuine or simply artifacts of the algorithms used, the clustering results are compared with those found in a control set. The control set contains pairs which satisfy the strong ties condition, but where the egos and alters are both non-movers. These non-mover egos are assigned dummy moving months which follow a distribution similar to that in the original dataset. In addition, we analyse the Spearman correlation coefficient for the unstandardised counts and fractions of the calls across different months for both the actual mover and dummy mover pairs.

\subsection*{Predicting post-move behaviour using pre-move behaviour as well as demographic and location information}

Next we predict the post-move behaviour using the users' demographic and location information and their pre-move behaviour. We specifically look at the following features: age and gender of the ego, age and gender difference between the ego and the alter, distance between the ego's home location and the alter's home location prior to and after the move, distance moved by the ego, and pre-move calling patterns. These pre-move calling patterns are the number and fraction of calls mentioned above, as well as a reciprocity measure for the calls, which is defined as follows:
\begin{equation}
\frac{c_\mathrm{ego \rightarrow alter} - c_\mathrm{alter \rightarrow ego}}{c_\mathrm{ego \leftrightarrow alter}}
\label{eq:recip_defn}
\end{equation}
\noindent where $c$ is the non-standardised number of calls, and the subscript refers to whether these are calls made by the ego to the alter, the alter to the ego, or the calls made between the ego and the alter irrespective of direction. The reciprocity measure is calculated per month and has a range of $[-1,1]$.

To compute for the distances related to the move, we have to find the locations of the ego and the alter before and after the move at a fine spatial resolution. We look for the home location of the egos and the alters one month before and two months after the estimated moving month (i.e., $t=-1$ and $t=2$) using the same procedure as outlined above, but at the city level rather than the province level. Since the actual move may have happened at $t=0$ or early $t=1$, the home locations at $t=-1$ and $t=2$ are more likely to reflect the true pre-move and post-move locations. We also note that since no time limit to night-time is considered in estimating the home location, this procedure may have picked up a city that is not the true home location, but perhaps a work or school location. However, we expect that the distance between the true home location and the inferred home location is small compared to the actual moving distance between the old home province and the new home province~\cite{song_limits_2010}.

In some cases (101 out of the 4,487 pairs), the home city location of the ego turns out not to be in its home province location. This can be explained by the aggregation method used: for example, suppose that the user is in the same province (e.g., province A) but in different cities (e.g., cities 1, 2, and 3, all in province A) for most of the time, but stays in one city in another province (e.g., city 4 in province B) for some time. If the number of days in city 4 is greater than the number of days in cities 1, 2, and 3 individually, the inferred city location can be city 4 even if the user spent the most time in province A. In such cases, we use the records taken in the inferred province home location and take the most common city location in this subset of records for the month.

From the clustering results, we find that we can aggregate the post-move behaviour as the mean in the post-move months. We focus on predicting (a) the post-move mean number and fraction of calls and (b) whether the number and fraction of calls decay after the move or not. We create a train-test set with an 80-20 split ($n_\mathrm{train}=3,583$, $n_\mathrm{test}=904$)
and use 5-fold cross validation when tuning the model parameters. Categorical features with $k$ levels were dummy-encoded into $k-1$ binary variables, and all features were normalised to mean 0 and standard deviation 1 using the train set. For the regression task (a), we use linear regression (ordinary least squares (OLS), ridge (Ridge), elastic net penalty (ElasticNet)), random forest regression (RF), $k$-nearest neighbour regression (KNN), and support vector regression (SVR) with the linear (SVR-lin), polynomial (SVR-poly), and radial basis function (SVR-RBF) kernels, while for the classification task (b), we use logistic regression with L2 penalty (LogReg), random forests (RF), AdaBoost, and support vector machines (SVM) with the linear (SVM-lin), polynomial (SVM-poly), and RBF (SVM-RBF) kernels. The models were chosen to give a good representation of the most commonly used linear and nonlinear models for regression and classification. We also note that neural networks were not considered due to the relatively limited size of the dataset. Models were implemented in Python using the \texttt{scikit-learn} module.

\section*{Results}
\label{sec:results}

\subsection*{General statistics}

\begin{figure}[!h]
    \includegraphics[width=\textwidth]{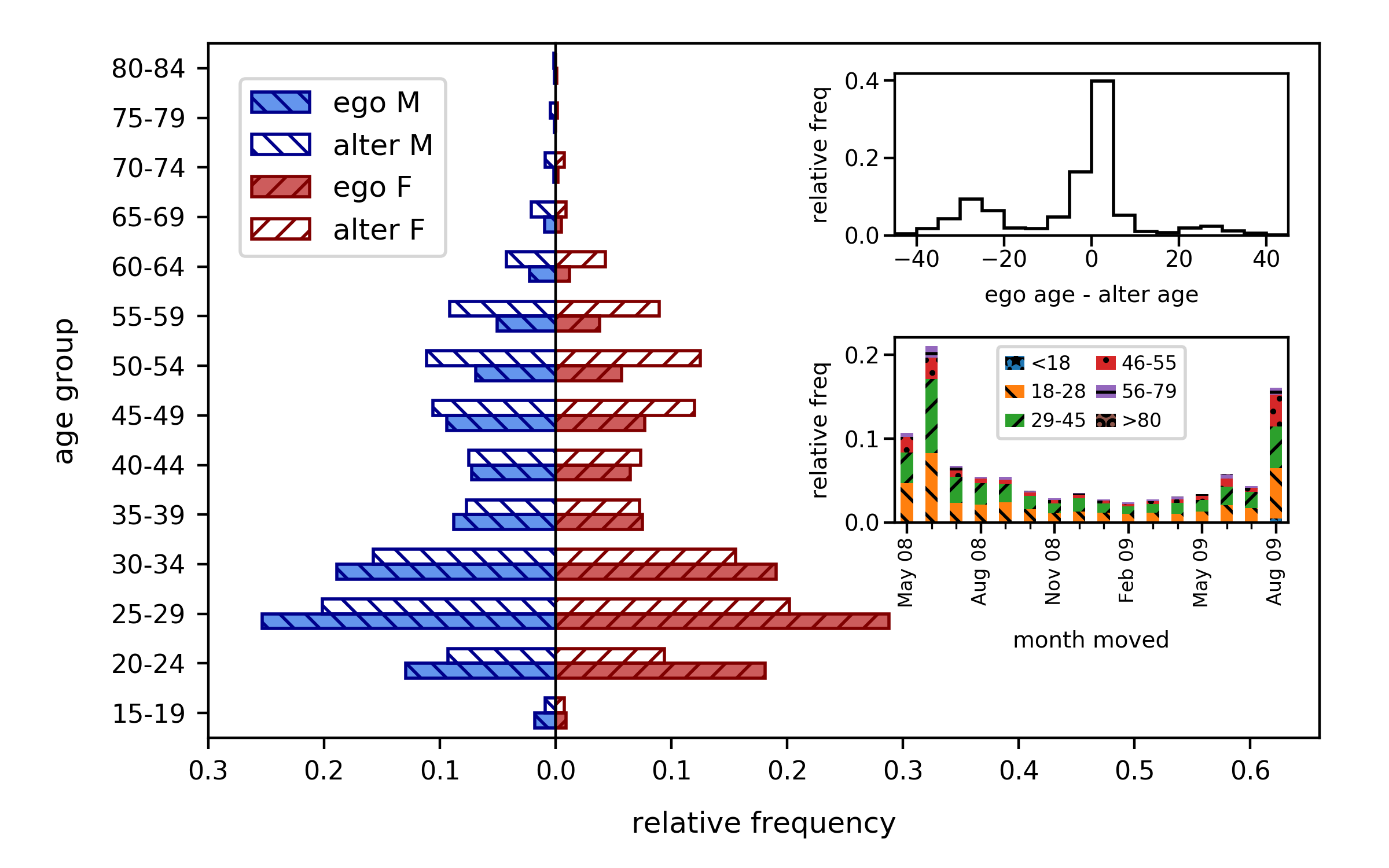}
    \caption{\csentence{Mobile phone user statistics.} The main figure shows the population pyramids for the egos (filled bars) and alters (unfilled bars) showing the relative frequencies for each gender and age group. The upper inset shows the relative frequency histogram of the age difference among the pairs considered, while the lower inset shows the relative frequency histograms of the egos' moving months, with the composition by age group indicated by the stacked bars. Due to the requirement that the ego and alter must be active for four months before and after the moving month, none of the egos in our analysis moved before May 2008 or after August 2009.}
    \label{fig:user_demog}
\end{figure}

The demographics of the users are shown in Figure~\ref{fig:user_demog}. Most of the mover egos are in the age group of 20--40 years old, while the alters are in two age groups, namely those in the same age group as the mover ego, with an age difference within 10 years, and the others in the age group of 20--40 years older. These likely correspond to the peers and parents, respectively, of the mover egos due to the age difference and the frequency and regularity of calls, as we had hypothesised in our previous work~\cite{fudolig_different_2020,fudolig_link-centric_2020,david-barrett_communication_2016}. We also note that the demographic profile of the full dataset is comparable with that of the census data, although users above 60 years of age are underrepresented due to a lack of mobile phone use among older individuals~\cite{ghosh_quantifying_2019}. Figure~\ref{fig:user_demog} also includes the distribution of the moving months as an inset; more details are included in the Supplementary Information (Figure S1).

\begin{figure}[!h]
    \includegraphics[width=\textwidth]{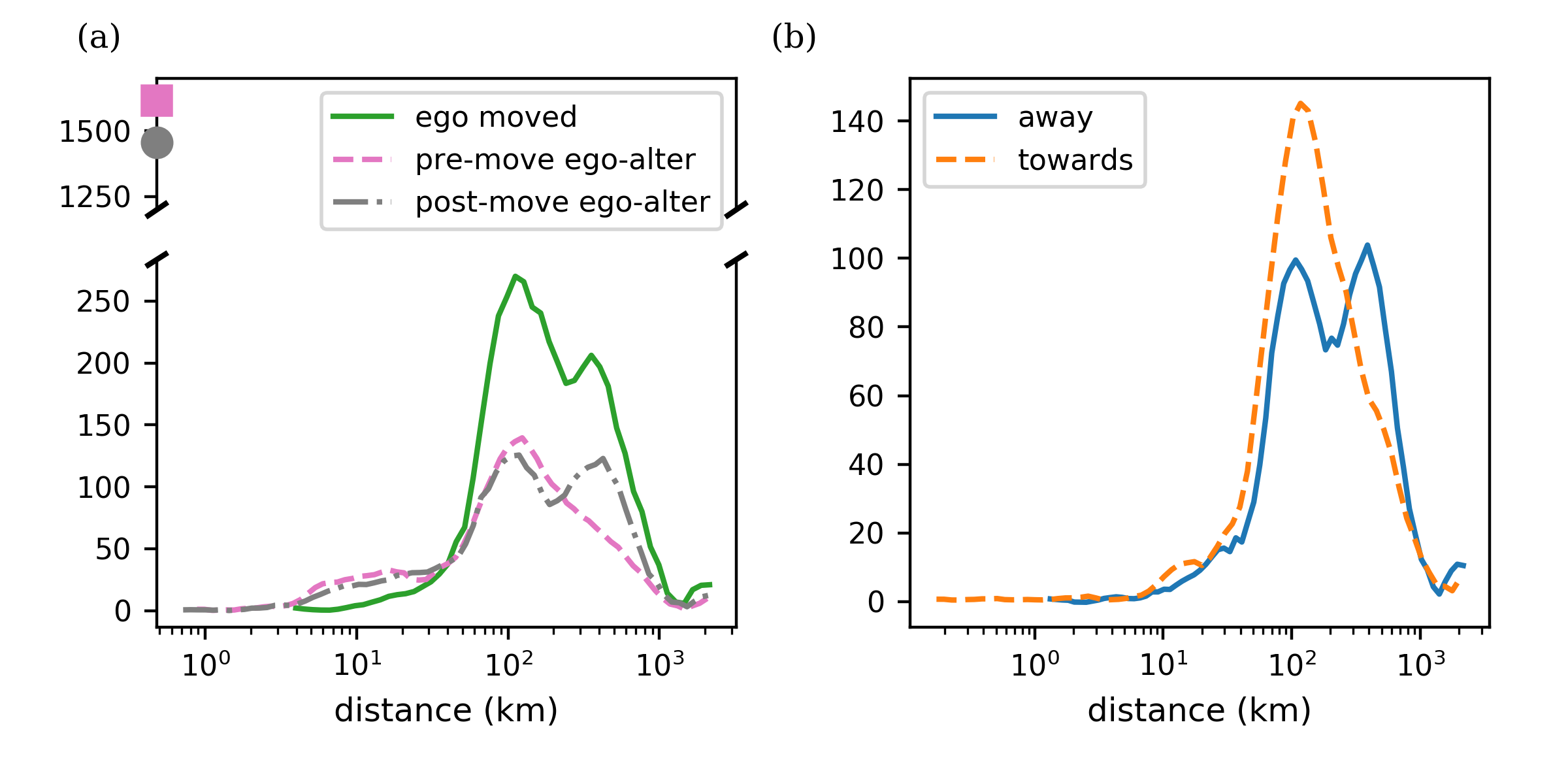}
    \caption{\csentence {Distances relevant to the migration.} The distributions of the distance moved by the ego (\texttt{ego moved}) and the pre- and post-move distances between the ego and the alter (\texttt{pre-move ego-alter} and \texttt{post-move ego-alter}) are superimposed in the left panel (a), with the filled square and the filled circle in the upper left corner indicating the number of pairs where the pre- and post-move distances between the ego and alter, respectively, are zero. The right panel (b) shows the distributions of the absolute difference of the distance between the ego and the alter before and after moving, with separate curves for egos moving away from the alter and egos moving towards the alter. All curves were obtained by using the Savitsky-Golay filter on the raw histograms.}
    \label{fig:gen_stat_distance}
\end{figure}

As expected, our method selects mostly egos who move relatively long distances: 94\% move at 50 km and above with the median moving distance being 168 km (Figure~\ref{fig:gen_stat_distance}(a)), although there are a number of cases where the ego moves a short distance crossing the provincial borders. We find that 47\% of the pairs were within 50 km of each other before the move and ended up moving away from each other, while in 43\% of pairs, egos moved closer to within 50 km of the alters' location. Around 5.6\% of the pairs continued to be within 50 km of each other before and after the move. 68\% of the pairs had the ego and the alter in the same city either before (36\%) or after (32\%) the move. Among the pairs which were initially in different locations before moving, the median pre-move ego-to-alter distance was 118 km; among pairs in different locations after moving, the median post-move ego-to-alter distance was 153 km  (Figure~\ref{fig:gen_stat_distance}(b)).

\subsection*{Clustering results}

We used \textit{k}-means clustering on the time series for each pair of both (1) the number of calls exchanged by the ego and the alter and (2) the fraction of the ego's calls that were associated with the alter. Each time series was standardised to have mean 0 and standard deviation 1, over the range $-4 \leq t \leq 4$, where $t=0$ corresponds to the moving month.

\begin{figure}[ht!]
    \centering
    \includegraphics[width=\textwidth]{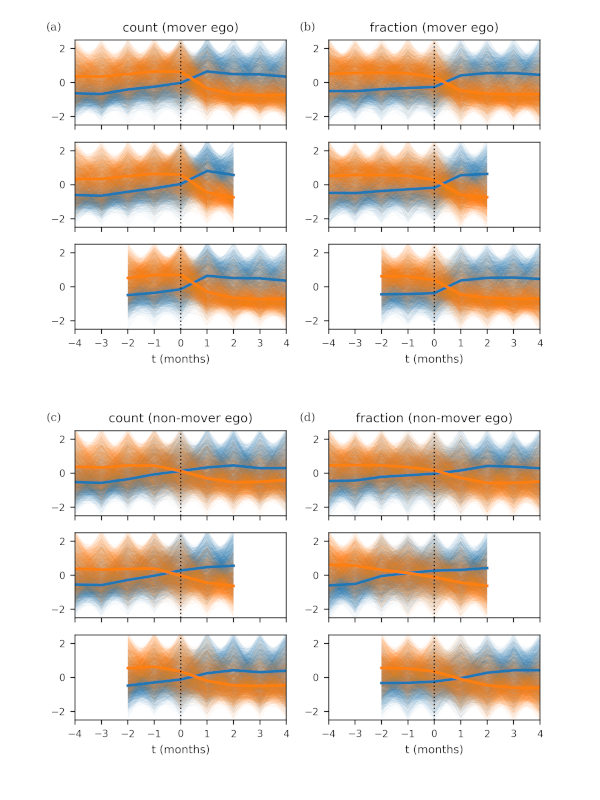}
    \caption{\csentence{Clusters found in the original vs. control datasets.} These plots show the clustering results on the standardised time series. Note that the rise or decay happens in various locations for the control dataset where the ego is a non-mover assigned a dummy moving month ((c) and (d)) but remains fixed (between $t=0$ and $t=1$) for the original dataset where the ego is a mover ((a) and (b)). A dotted line at $t=0$ is included for guiding the eye.}
    \label{fig:mover_vs_non-mover}
\end{figure}

We find that the silhouette index, Davies-Bouldin score, and the Jaccard bootstrap similarity index~\cite{yu_bootstrapping_2019} all gave $k=2$ as the optimum number of clusters for the fraction as well as for the number of calls (Supplementary Information Figure S2). The resulting cluster prototypes obtained by averaging over all cluster members show two typical behaviours of pairs: those where the quantity of interest increased and those where it decreased. In both cases, the change occurs at around $t=0$ as shown in Figure~\ref{fig:mover_vs_non-mover}(a,~b). Similar results were also obtained using Ward's method.

To ensure that this result is not an artifact of the data preprocessing and the clustering algorithm, we compare our results to a control dataset of the same size where both the egos and alters are non-movers, but the ego is assigned a dummy month with a distribution similar to that in the original dataset. The results are shown in Figure~\ref{fig:mover_vs_non-mover}(c,~d). We note that both the original and the control dataset from $t\in[-4,4]$ showed two cluster prototypes, one where the quantity of interest increases and the other decreases, and the change happens at around $t=0$. As there is no real meaning to $t=0$ for the control dataset, the month of changes ($t=0$) for this dataset could be a trivial result in contrast to that for the original dataset. Due to the standardisation of time series to mean $0$, they will cross the zero axis but at arbitrary times as the moving months have dummy values. Thus, averaging those time series results in crossing the zero axis in the middle of the interval $[-4,4]$, i.e., at around $t=0$. By the same reason, we expect for the control dataset that if we truncate the standardised time series, the averaged time series will cross the zero axis in the middle of the truncated interval. This explains why in the control dataset, a $k=2$ clustering yields rising and decaying prototypes with the rise or decay occurring in the middle of the interval, shifting in location depending on how the interval is truncated. In contrast, for the original dataset, we find that even when performing \textit{k}-means clustering on the truncated time series, the rise or decay happens at the same point in time around $t=0$. Thus, the rising and decaying prototypes in the original dataset can be attributed to the significance of $t=0$.

\begin{figure}[!h]
    \includegraphics[width=\textwidth]{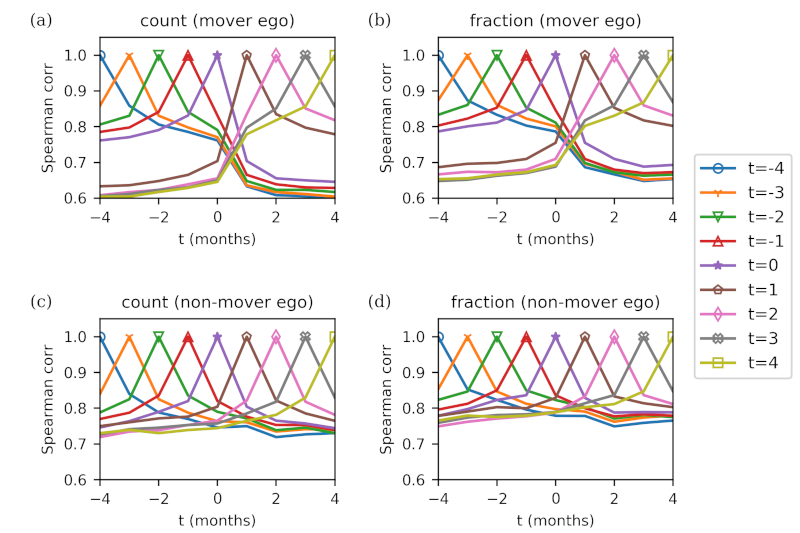}
    \caption{\csentence{Spearman correlation coefficients for calling frequency quantities between different months.} These plots show the Spearman correlation coefficients between the unstandardised number or fraction of calls in one month and another month for pairs with mover egos ((a) and (b)) and pairs with non-mover egos ((c) and (d)). Each line corresponds to a particular month, and shows the correlation coefficients between that month and all values of $t$. The correlation coefficient between the particular month and itself simply has a value of $1$.
    }
    \label{fig:corrcoeff_mover_non-mover}
\end{figure}

The significance of the estimated moving month $t=0$ can also be seen by measuring the correlation of the number or fraction of calls between different months. Precisely, for each value of $t$, we take the Spearman correlation coefficient between the unstandardised quantity (count or fraction) in that month and the same quantity in another month. Among the pairs where the ego moved, there is a high positive correlation between pre-move months ($t\in[-4,-1]$) as well as between post-move months ($t\in[1,4]$), and the values of the correlation coefficients between pre-move months are similar to those between post-move months. However, the correlation between a pre-move and a post-move month is markedly lower. Although it is expected that the correlation is higher for consecutive months compared to non-consecutive months, the sharp difference in the drop in correlation coefficients at $t=0$ is observed among pairs with mover egos but not among pairs with non-mover egos (Figure~\ref{fig:corrcoeff_mover_non-mover}). When we examine the correlation coefficients for each cluster separately, we find that the correlations across all months remain high (Supplementary Information Figure S3), with Spearman's $\rho\geq0.7$ for both the number and the fraction of calls. Thus, the drop in the correlation coefficient observed in the aggregate population (no separation by cluster) is likely due to the difference in the post-move behaviour in the two clusters found.

We also note that the vast majority of pairs behave consistently with their corresponding cluster prototype, with a 97\% agreement between the cluster labels (rise or decay) and the actual changes (increase or decrease) in the number and fraction of calls. For pairs where the cluster label does not agree with an actual rise or decay in the number or fraction of calls, the difference between the pre-move and post-move values is much smaller than those observed in properly labelled pairs. Further, there is a considerable change between pre-move and post-move values. The median change in the number of calls after moving is 42\%, while the median change in the fraction of calls after moving is 34\%. Separating the sample into the rise and decay clusters give similar results for each cluster; detailed results are summarised in the Supplementary Information (Supplementary Information Table S2). Taken together, these bolster our finding that pairs with a mover ego and a non-mover ego are best clustered into two major groups, one where communication rises and another where it decays.

\subsection*{Predicting post-move calling frequency volume from pre-move calling behaviour as well as demographic and location information}

The moderate correlation between the pre-move and post-move values shown in Figure~\ref{fig:corrcoeff_mover_non-mover} suggests that we may be able to predict post-move behaviour if the pre-move behaviour is known. Therefore, we will discuss a number of predictive models and explore the extent to which the pre-move behaviour coupled with demographic and location information can predict the post-move call frequency.

We first examine whether we can predict the number of calls and the fraction of calls made and received by the ego that are associated with the alter after the move. Since the number and fraction of calls are highly correlated between the pre-move months as well as between the post-move months, we take the means of the values across the pre-move months and across the post-move months and use these means to characterize the pre-move and post-move behaviour, respectively. We predict the post-move means of the number and fraction of calls by using the following as predictors:

\begin{enumerate}[(i)]
    \item the mean of the number of calls between the ego and the alter across the pre-move months, also denoted for brevity as the ``pre-move count'' (\texttt{count\_pre})
    \item the mean of the fraction of calls between the ego and the alter across the pre-move months, also denoted for brevity as the ``pre-move fraction'' (\texttt{frac\_pre})
    \item age of the mover ego (\texttt{age\_ego})
    \item age difference between the ego and the alter, i.e., ego's age minus alter's age 
    (\texttt{age\_diff}) 
    \item gender of the mover ego (categorical, \texttt{gender\_ego})
    \item gender difference between the ego and the alter (same or opposite gender, categorical, \texttt{gender\_diff})
    \item distance moved by the ego in kilometers (\texttt{distance\_move})
    \item pre-move distance between the ego and the alter in kilometers (\texttt{distance\_ea\_pre})
    \item direction of move (towards or away from the alter, categorical, \texttt{direction\_move})
    \item pre-move mean reciprocity measure in Equation~\ref{eq:recip_defn} (\texttt{recip\_pre})
\end{enumerate}

The absolute value of the difference between the pre-move distance and the post-move distance between the ego and the alter was also considered, but was discarded due to its very strong correlation with the distance moved by the ego (Pearson's $\rho=0.96$). Most of the predictors used were not correlated with each other ($\lvert \rho \rvert<0.15$), except for \texttt{age\_diff} and \texttt{age\_ego} ($\rho=0.51$), \texttt{distance\_ea\_pre} and \texttt{direction\_move} (point bi-serial correlation coefficient $\lvert r_{pb} \rvert =0.50$), and \texttt{frac\_pre} and \texttt{count\_pre} ($\rho=0.72$). The log transformation was considered for particular variables with considerable skew (the mean number of calls and distances), and the logit transform was also considered for the mean fraction of calls. For the log and logit transformation of zero values, we substitute the transform of a sufficiently small value (e.g., for pre-move distance in kilometers and for the number of calls, we use $0.1$ instead of $0$; for the fraction of calls, we use $0.001$ instead of $0$). Various combinations of transformed and untransformed variables were compared for performance using the $R^2$ score and the mean squared error (MSE) computed on the test set. In cases where the response variable is transformed, these scores are computed by first back-transforming the predicted variable into its original scale, allowing us to compare models with transformed and untransformed response variables. Categorical predictors (gender and direction of move) were dummy encoded, and all predictors were standardised prior to inclusion in the model.

\subsubsection*{Number of ego's calls associated with the alter}

\begin{figure}[!h]
    \includegraphics[width=\textwidth]{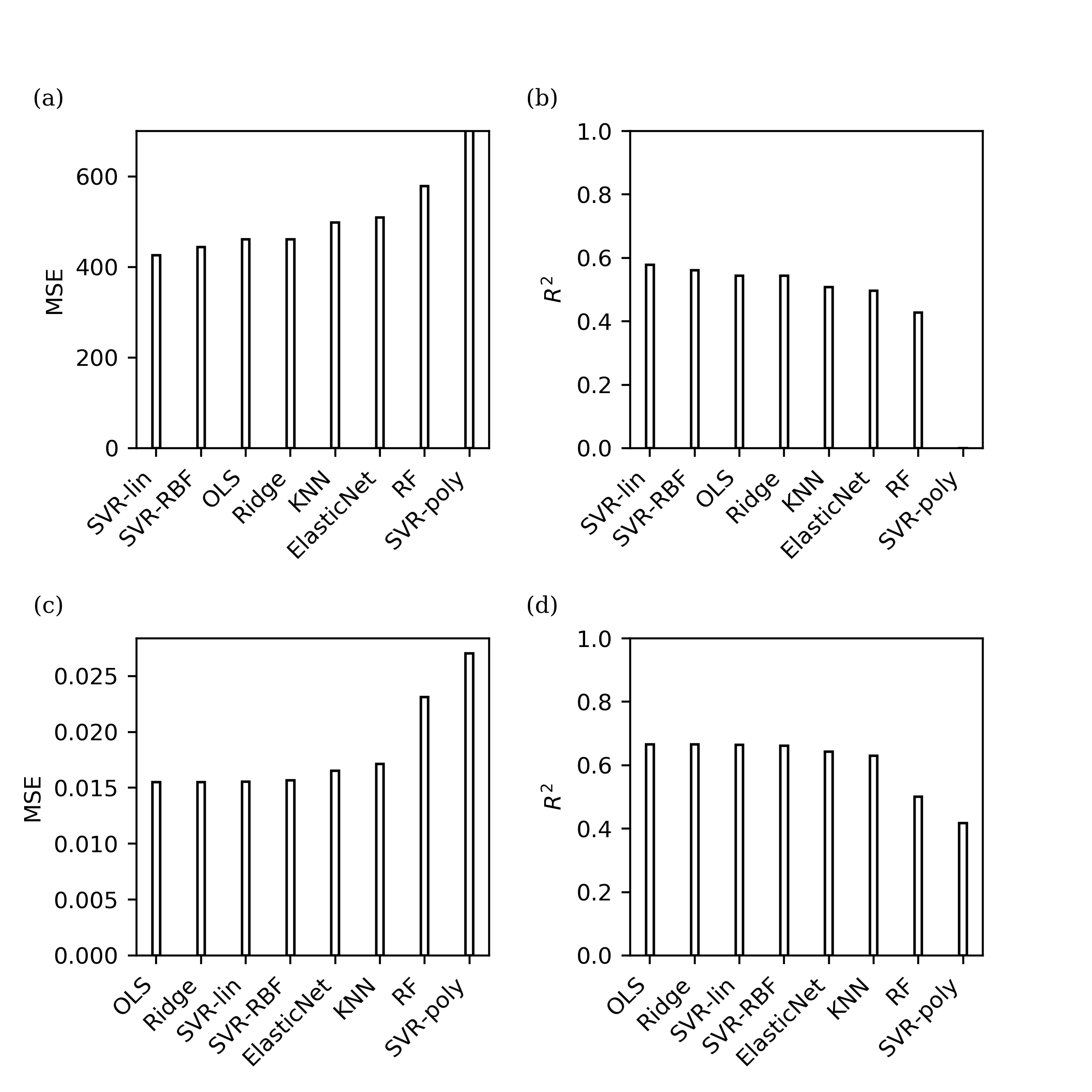}
    \caption{\csentence{Regression metrics.} These are the MSE and $R^2$ values for predicting the post-move means of the number of calls ((a) and (b)) and the fraction of calls ((c) and (d)) using the log-transformed pre-move counts ((a) and (b)) and the pre-move fractions ((c) and (d)). The $R^2$ for the SVR with a polynomial kernel in predicting the post-move counts is negative, while the corresponding MSE ($1.3\times10^7$) is much higher than the MSE of the other models.
    }
    \label{fig:regression_metrics}
\end{figure}

\begin{figure}[!h]
    \includegraphics[width=\textwidth]{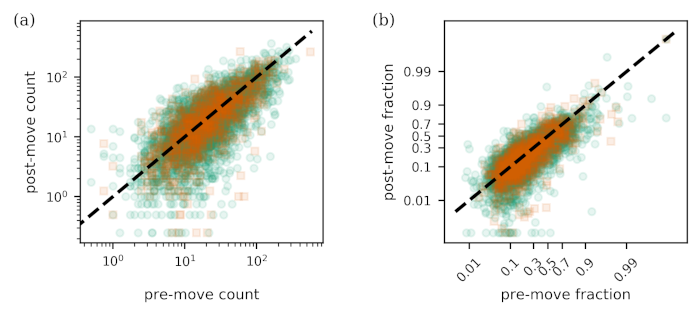}
    \caption{\csentence{Post- vs pre-move number and fraction of calls.} These plots show the post- vs. pre-move means of the (a) number of calls (log scale) and (b) fraction of calls (logit scale). Green circles refer to the train set and orange squares refer to the test set, while the dashed line corresponds to the $y=x$ line. Note the clear linear trend and the considerable variance about the $y=x$ line, indicating that although the pre-move value alone may give a rough estimate of the post-move value, it may not be enough to determine if the post-move value will be higher or lower than the pre-move value (above or below $y=x$).
    }
    \label{fig:post_vs_pre_frac_count}
\end{figure}

In predicting the number of calls between the ego and the alter, the best result, both in terms of the test $R^2$ value (0.58) and the test MSE (426.0), was given by the linear SVR with only one predictor, i.e., the pre-move count, and with both the response and the predictor log-transformed. The SVR with an RBF kernel, ordinary least squares regression, and ridge regression have similar performance, while the SVR with a polynomial kernel performed extremely poorly, as depicted in Figure~\ref{fig:regression_metrics}(a,~b). The linear trend between the pre-move and the post-move counts can also be observed in Figure~\ref{fig:post_vs_pre_frac_count}(a). We also note that the best performing models for the complete sets of predictors give comparable performance ($R^2$ from 0.55 to 0.58, MSE from 429.5 to 452.2). 

These indicate that while the pre-move count explains much of the variance in the post-move count, the remaining variance cannot be explained sufficiently by the demographic information of the pair, the distance moved by the ego, the distance between the ego and the alter, and the reciprocity of calls between the ego and the alter. On the other hand, the pre-move fraction, being moderately positively correlated with the number of pre-move count (Spearman's $\rho=0.72$), seems to be a redundant predictor.

\subsubsection*{Fraction of ego's calls associated with the alter}

The results for predicting the post-move mean of the fraction of calls of the ego associated with the alter are similar to those found for predicting the post-move mean of the number of calls. The best model is obtained using ordinary least squares regression with only one predictor, i.e., the pre-move fraction ($R^2=0.67$, $\textrm{MSE}=0.015$). Linear models, as well as the SVR with an RBF kernel, exhibit similar performance (Figure~\ref{fig:regression_metrics}(c,~d)). The linear trend between the pre-move and the post-move fraction can also be observed in Figure~\ref{fig:post_vs_pre_frac_count}(b). We also note that the best performing models on complete sets of predictors yielded comparable performance ($R^2$ from 0.64 to 0.66, MSE from 0.016 to 0.017). Again, these indicate that though the pre-move behaviour may be used to predict the post-move behaviour, the demographic and location information do not improve much the prediction performance.

\subsection*{Predicting rise or decay in calling frequency from pre-move calling behaviour as well as demographic and location information}

We have observed that the post-move calling frequency volume (in terms of both the number and fraction of calls) can be predicted, to some extent, using the pre-move behaviour, whereas demographic and location information do not significantly improve performance. We now attempt to see if we can at least predict the \textit{direction} of change in the calling frequency by predicting whether the post-move calling frequency (mean number or fraction of calls) is lower than the pre-move value. These roughly correspond to the rise and decay clusters from the clustering analysis, but instead of using the cluster labels as ground truth, we use the actual pre- and post-move means to determine the rise or decay. We also note that there are a few cases where the pre- and post-move means are the same, though they comprise a very small proportion of the pairs (0.96\% for the number of calls and 0.04\% for the fraction of calls). These pairs are lumped together with those whose calling frequency increased post-move. Though this group is more aptly named ``non-decay'' cluster, we keep the term ``rise'' for simplicity, as the majority of this group behave in this way. The resulting train set is mildly imbalanced with the decay-rise composition of 57\%-43\% and 62\%-38\% distribution for the number and fraction of calls, respectively, and we remedy the imbalance by using balanced class weights in the classification algorithms as implemented in the Python module \texttt{scikit-learn}.

The basic procedure is the same as in the regression task, but we predict a binary output using classification models. We consider two metrics: the total accuracy and the average of the recall for each class (rise and decay), with the latter equivalent to the average of the accuracy scores for each class considered separately.

\subsubsection*{Number of ego's calls associated with the alter}

\begin{figure}[!h]
    \includegraphics[width=\textwidth]{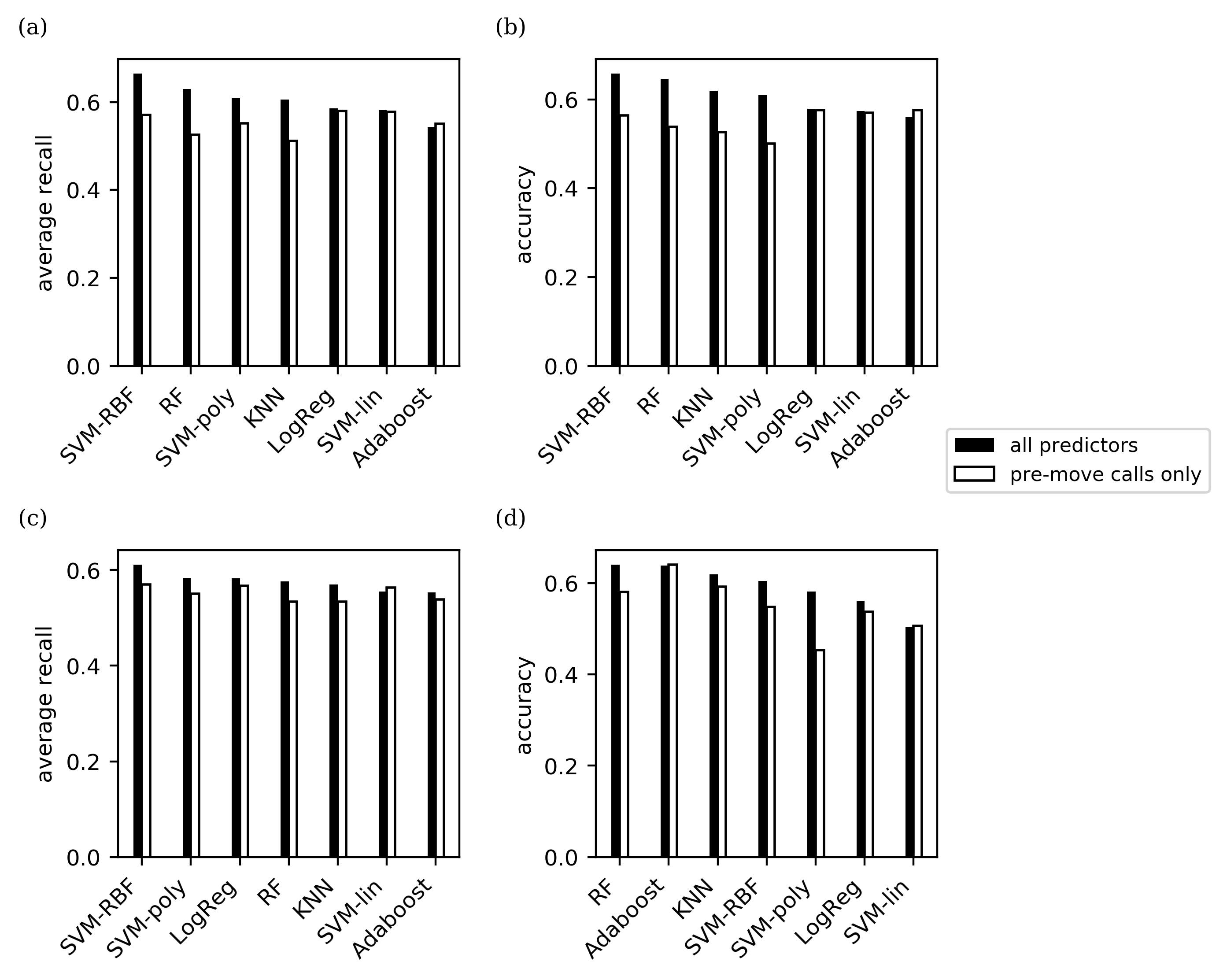}
    \caption{\csentence{Classification metrics.} These are the average recall and total accuracy values for predicting whether the post-move means of the number ((a) and (b)) and the fraction of calls ((c) and (d)) decay or not. The dark bars show the average recall and total accuracy for the model using all predictors, while the white bars show the same metrics for the model using only the pre-move means of the number of calls and the pre-move fraction (untransformed fraction and log-transformed count).
    }
    \label{fig:classification_metrics}
\end{figure}

The SVM-RBF classifier gives the best performance both in terms of the total accuracy and average recall (Figure~\ref{fig:classification_metrics}(a,~b)). Using all predictors, where the pre-move counts and distances are log-transformed, gives a total accuracy and average recall of 0.66, with the recall for the rise class at 0.71 and the decay class at 0.62. The precision for the rise and decay classes are 0.58 and 0.74, respectively.

\begin{figure}[!h]
    \includegraphics[width=\textwidth]{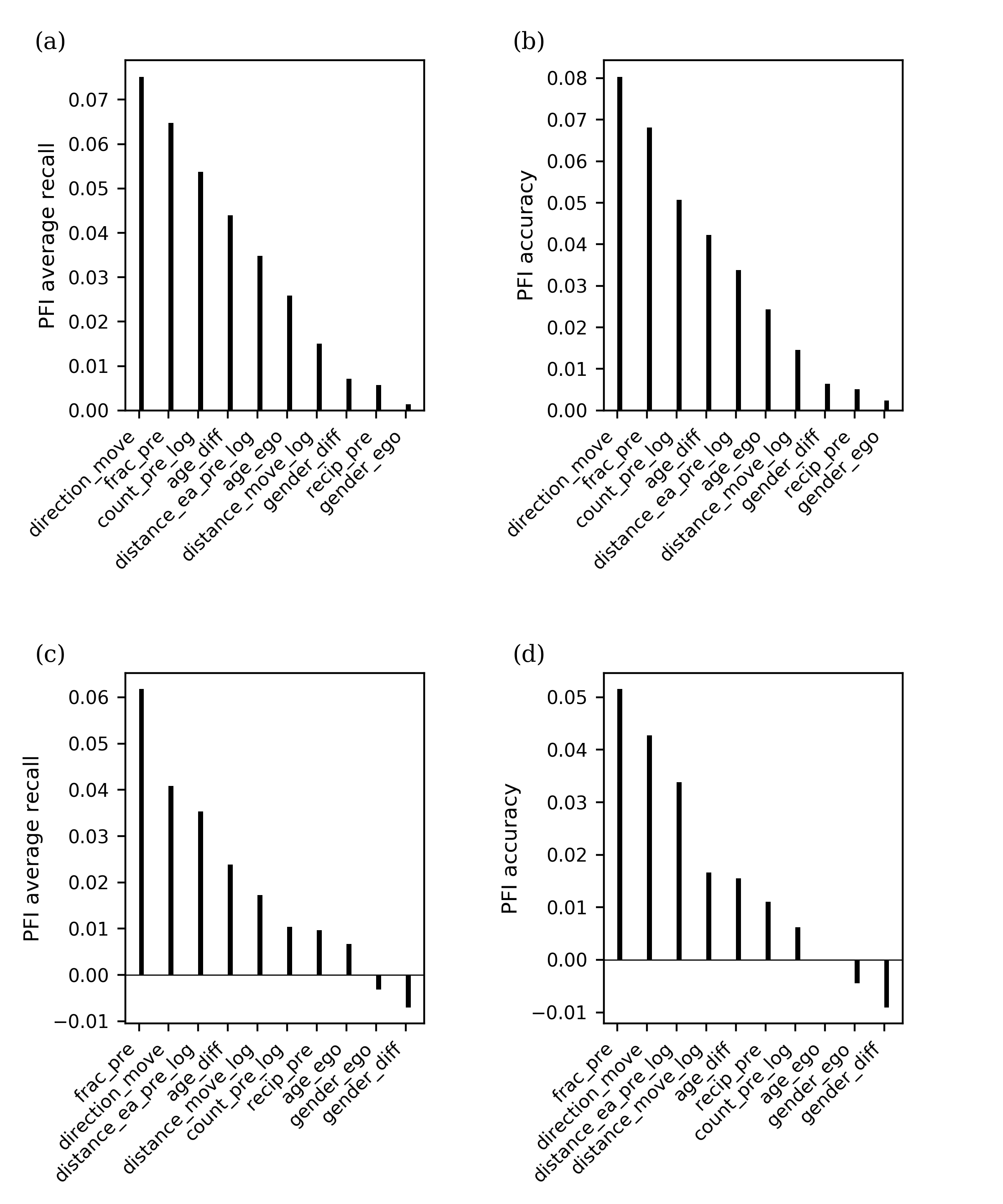}
    \caption{\csentence{Permutation feature importance for rise/decay classification.} These show the permutation feature importance of each predictor used in the best model for predicting rise/decay in the mean number of calls ((a) and (b)) and the mean fraction of calls ((c) and (d)). The more positive the permutation feature importance for a particular predictor is, the more shuffling the predictor values negatively affects the model performance.
    }
    \label{fig:classification_perm_impt}
\end{figure}

To obtain the most relevant predictors, we calculate the permutation feature importance as in Ref.~\cite{breiman_random_2001}, which is the difference from the baseline metric when the values of the predictor are shuffled across all pairs. The more positive the permutation feature importance, the better the baseline metric is over the metric with the shuffled predictor (for metrics that are meant to be minimised, the sign is reversed so that a more positive permutation feature importance still corresponds to a decline in metric quality). For our implementation, we take the mean of the difference obtained from five different permuted sets of the test set. In contrast to the regression task, we find that the demographic and location information are also relevant predictors. The permutation feature importances are not heavily skewed towards only one predictor but are within the same order of magnitude across most predictors. For both the average recall and the total accuracy, the most relevant predictors in the best performing model are the direction of the move (away or towards the alter), the pre-move fraction, the pre-move count, and the age difference, as shown in Figure~\ref{fig:classification_perm_impt}(a,~b).

In order to further illustrate the relevance of the demographic and location information in the prediction performance, we look at the performance of the models but with only the pre-move fraction and the (log-transformed) pre-move count as predictors. Unlike in the regression task, removing the demographic and location information results in noticeably poorer performance as shown in Figure~\ref{fig:classification_metrics}(a,~b), except in the case of the linear models and AdaBoost (note that AdaBoost tends to perform very well in predicting the decay class and very poorly in the rise class, and thus acts like a dummy majority classifier). This indicates that the relationship of demographics and location to post-move calling behaviour is likely to be nonlinear.

\subsubsection*{Fraction of ego's calls associated with the alter}

Although the best accuracy scores are obtained using the tree-based classifiers (i.e., random forests and AdaBoost), the resulting models behave like dummy majority classifiers, with a very high recall for the decay class (0.83--0.89) and a very low recall for the rise class (0.21--0.31). In contrast, the SVM-RBF classifier gives a higher average recall with comparable recalls for each class (0.59--0.60 for the decay class, 0.58--0.64 for the rise class). Overall, the best performing model is the SVM-RBF classifier with the pre-move count and the distances log-transformed, with a total accuracy score and average recall score of 0.61, see Figure~\ref{fig:classification_metrics}(c,~d). The recall for the rise and decay classes are 0.64 and 0.59, respectively, while the corresponding values for the precision are 0.48 and 0.73.

The most relevant predictors for this model are the pre-move fraction, the direction of the move (away or towards the alter), and the pre-move distance between the ego and the alter (Figure~\ref{fig:classification_perm_impt}(c,~d)). Using only the pre-move fraction and the (log-transformed) pre-move count gives only slightly poorer predictive performance than using all predictors, as shown in Figure~\ref{fig:classification_metrics}(c,~d).

\section*{Discussion and concluding remarks}

In contrast to surveys most commonly used in migration studies, mobile phone CDRs give us a very detailed quantitative view on how the mobile communication patterns of movers change, while being unaffected by the recall bias and characterised by high spatiotemporal resolution. We were primarily interested in how a long-distance residential move would affect the mobile communication patterns between a mover ego and a non-mover alter that frequently and regularly communicated with the ego prior to the move. In particular, we wanted to investigate the role of the demographic and location information of the ego-alter pair on these changes, with the demographic information serving as a proxy for the relationship between a consistently communicating pair of users.

We focused on two quantities that characterise the mobile communication patterns: (a) the number of calls between the mover ego and the non-mover alter, and (b) the fraction of calls made and received by the mover ego that are associated with the non-mover alter. The number of calls serves as a measure of communication volume independent of the other alters of the ego, while the fraction of calls is a proxy for the relative importance of the alter to the ego. Using clustering analysis, we have found that a change in the communication patterns happens shortly after the move, generally speaking either rising or decaying. As there is a high correlation both within the pre-move months and within the post-move months, but a lower correlation between the pre-move and post-move months, the extent of this change is limited mostly to the moving month, with the communication stabilising shortly after. Interestingly, very few pairs cut communication entirely. At four months after moving, only 3.5\% of the close pairs we examined have no calls or SMS. A preliminary investigation reveals that the pairs that stopped communicating are disproportionately composed of young peers, and we aim to study this phenomenon in more detail in the future.

We also find that the post-move means of the number and fraction of calls can be predicted to some extent by the corresponding pre-move values alone, although a large proportion of the variance remains unexplained even when demographics and the distances moved are taken into account. The relationship between the pre- and post-move calling frequency volumes shows a linear trend (Figure~\ref{fig:post_vs_pre_frac_count}), with the pre-move values likely resulting in similar post-move values, consistent with the findings using other datasets~\cite{hong_characterization_2019}. In predicting the rise or decay in the number and fraction of calls, the demographic and location information provide more information to nonlinear models, yielding better performance than if only the pre-move means of the number and fraction of calls had been used as predictors. Thus, whereas a higher frequency of pre-move calls generally leads to a higher frequency of post-move calls, demographics and migrating distances appear to have a more complicated nonlinear effect on the post-move calling behaviour. We also note that in predicting the rise or decay of calling frequency, the most relevant predictors other than the pre-move calling behaviour involve the direction of migration, migrating distances, and the age difference of the pair, while the genders of the ego and the alter have little effect on the prediction performance. As in our dataset, the age difference mostly falls into one of the two groups, one corresponding to the age differences among peers (0--10 years) and the other to parent-child or similar relationships (20--40 years), our results suggest that the type of relationship between pairs affects post-move communication, which is consistent with earlier sociological findings~\cite{wellman_decade_1997}.

Although the predictive performance of our classifiers is better than random, both the average recall and the accuracy are less than 70\%. We can compare this to the \textit{Bayes accuracy}, the theoretical upper bound of the accuracy obtained by any classifier for a given sample. The Bayes accuracy itself cannot be calculated without the knowledge of the underlying distributions, but its bounds can be estimated. The estimate for the upper bound of the Bayes accuracy obtained from the 1-nearest neighbour classifier~\cite{tumer_estimating_1996} is 70\% for both classification tasks. As the training set size is not very large, we expect some bias in this estimate, but as the dimensionality of the prediction problem is not very high, we do not expect the actual upper bound to be considerably higher~\cite{fukunaga_bias_1987}. Thus, even though we did not exhaust all possible models and algorithms, it is more likely that the lack of predictive power can be attributed more to the inherent overlap in the feature space of the two classes (decay and rise) than to the insufficiency of the models used. In other words, the post-move behaviour of the pairs where the ego moved but the alter did not, may also be influenced by other variables that we did not consider in our models.

A limitation of using our dataset to study mobility is that a user's location is recorded only when the user makes a call or sends an SMS and when the cell tower used has known geographical coordinates. We have worked around this limitation through a combination of approaches. First, we increased the number of viable samples by not introducing a time limit in exchange for reducing the spatial resolution to the province level. Second, we obtained the home location by taking the majority province location at two levels: we first took the most common location per day, and then among these locations, the most common location per month. This prevents days with an unusually high number of calls to significantly influence the inferred home location. Lastly, we require that the user only moves once and that the home location stays stable over a certain period (i.e., the egos we considered were in the same home location for at least four months), making it more probable that the user indeed moved and not commuted between the inferred home locations. We also note that using the night-time calls and SMS to infer the home location results in more users with unknown home locations, but that if unknown home locations are excluded, the trajectories obtained using the all-day and the night-time calls and SMS are consistent for the majority of the movers and non-movers. By imposing these criteria to identify movers, we find that most of the mover egos that we included in our analysis moved long distances (Figure~\ref{fig:gen_stat_distance}), which require a lot of time or financial costs for daily commutes by European standards~\cite{guirao_labour_2018}, indicating that the move was likely a true residential move instead of simply a change in workplace. Our clustering results also bolster the relevance of the moving month in the call frequency time series, indicating that the movers and their moving month were likely correctly inferred.

We also note that while the size of the sample analysed is small, it is reasonable given the migration statistics of the country studied (see the Methodology section for details). Further, we emphasise that the highly specific nature of our target population, long-term migrants with close alters who did not move, does not reduce their relevance, as migration and its effects on social interactions have been of interest to researchers for several years. The use of CDRs with demographic and location information has paved the way to study this topic at a much larger scale and with higher spatiotemporal resolution. However, CDR datasets have inherent sampling biases as well. In our case, the data was gathered from only a single service provider (albeit with a significant market share) in a single European country, and the user locations are known only when they make a call or an SMS using a cell tower with a known location. It is also worth noting that these issues of sampling bias are common across most, if not all, CDR datasets. CDR datasets from other countries or service providers would facilitate to further test our approach and to investigate the interplay between the residential mobility and communication patterns of a wider class of ego-alter pairs.

In summary, our study demonstrates the predictability of post-move mobile communication based on pre-move mobile communication, migrating distances, and demographics of communicating individuals. To our knowledge, our work is the first of its kind in investigating the interplay between these factors from a quantitative and predictive standpoint. The usefulness of digital datasets of long term mobile communication is clearly evidenced in providing estimates with high accuracy and spatiotemporal resolution. Our study is also different from previous studies involving CDRs~\cite{phithakkitnukoon_out_2011,song_limits_2010,lu_predictability_2012,simini_universal_2012} as it focuses not just on mobility but also on both tie strength and demographics. Here, we restricted our attention to pairwise communication and, in particular, strong ties between the ego and the alter. In the future, using a similar methodology we intend to broaden the scope and examine the factors possibly influencing individual choices in migration~\cite{ghosh_migration_2019,blumenstock_migration_2019,buchel_calling_2020}. Further, whereas our current study deals with predicting post-move calling behavior from demographic and location information as well as pre-move calling behavior, an interesting direction for future work is to look at how different types of relationships, as can be inferred from a pair's demographic information, vary in terms of how the pair's calling behavior changes after a residential move.


\begin{backmatter}

\section*{Availability of data and materials}
The datasets analysed in the current study are available in a public online repository, \url{https://doi.org/10.6084/m9.figshare.12874028.v1}.

\section*{Competing interests}
  The authors declare that they have no competing interests.

\section*{Author's contributions}

All authors participated in the conceptualisation of the project. MIDF and DM curated the data, while MIDF developed the methodology and performed the data analysis. HHJ and KK supervised the project. MIDF wrote the first draft of the manuscript, while all authors contributed to its finalisation.

\section*{Acknowledgements}
We acknowledge the computational resources provided by the Aalto Science-IT project. We also thank Miguel Antonio Fudolig and Prof. Francisco de los Reyes for their invaluable inputs.

\section*{Funding}
DM, KB, and KK acknowledge support from EU HORIZON 2020 INFRAIA-1-2014-2015 program project (SoBigData) No. 654024 and INFRAIA-2019-1 (SoBigData++) No. 871042. KK also acknowledges the Visiting Fellowship at The Alan Turing Institute, UK. HHJ acknowledges financial support by Basic Science Research Program through the National Research Foundation of Korea (NRF) grant funded by the Ministry of Education (NRF-2018R1D1A1A09081919). The funders had no role in study design, data collection and analysis, decision to publish, or preparation of the manuscript.
  

\bibliographystyle{bmc-mathphys} 

\section*{Additional Files}
  \subsection*{Additional file 1 --- Supplementary Information}
    PDF file containing supplementary figures and tables


\end{backmatter}
\end{document}